\documentclass[a4paper,reqno]{amsart}
\usepackage{amsmath,amsthm,amssymb,amsfonts,xspace}
\usepackage[T1]{fontenc}
\usepackage[utf8]{inputenc}
\usepackage[english]{babel}
\usepackage{amsxtra}
\usepackage{enumerate}
\usepackage{verbatim}
\usepackage{color}
\usepackage[mathscr]{eucal}

\usepackage[bookmarksnumbered,colorlinks]{hyperref}
\usepackage[colorinlistoftodos,italian]{todonotes}
\usepackage[normalem]{ulem}
\usepackage{tensor}

\newcommand{\R}{\mathbb R}

\newtheorem{Proposition}{Proposition}

\theoremstyle{remark}
\newtheorem{rmk}{Remark}
\theoremstyle{definition}

\def\bal#1\eal{\begin{align}#1\end{align}}              
\def\baln#1\ealn{\begin{align*}#1\end{align*}}          
\def\bml#1\eml{\begin{multline}#1\end{multline}}        
\def\bmln#1\emln{\begin{multline*}#1\end{multline*}}  
\def\bga#1\ega{\begin{gather}#1\end{gather}}
\def\bgan#1\egan{\begin{gather*}#1\end{gather*}}

\newcommand{\beq}{\begin{equation}}
\newcommand{\eeq}{\end{equation}}
\newcommand{\bere}{\begin{rmk}}
\newcommand{\ere}{\end{rmk}}
\newcommand{\bpr}{\begin{Proposition}}
\newcommand{\epr}{\end{Proposition}}

\begin{document}

\title[The Sagnac effect as a Finslerian effect]{A note on the Sagnac effect in general relativity as a Finslerian effect}
\author[E. Caponio]{Erasmo Caponio}
\address{Department of Mechanics, Mathematics and Management, \hfill\break\indent
	Politecnico di Bari, Via Orabona 4, 70125, Bari, Italy\vspace{12pt}}
\email{erasmo.caponio@poliba.it}
\author[A. Masiello]{Antonio Masiello}
\email{antonio.masiello@poliba.it}
\thanks{E.C. and A.M. are partially supported by  PRIN 2017JPCAPN {\em Qualitative and quantitative aspects of nonlinear PDEs.}}


\begin{abstract}
	The geometry of the Sagnac effect in a stationary region of a spacetime  is reviewed   with the aim of emphasizing the role of asymmetry of   a  Finsler metric defined on a spacelike hypersurface associated to a stationary splitting and related to  future-pointing null geodesics of the spacetime. We show  also that an analogous asymmetry comes into play in the Sagnac effect for  timelike geodesics.

\end{abstract}

\maketitle

\section{Introduction}
A light beam constrained to follow  a closed path on a rotating system takes different times to reach a detector according to its travel direction. This is the Sagnac effect which allows  a non-inertial observer to compute the angular speed of the rotating  system and which is at the base of corrections adopted in synchronization of atomic clocks in GPS (see e.g. \cite{Ashby03}).

Even though it has been the subject of a longstanding quarrel  about the validity of special relativity (see, for example, \cite{Seller04}),  P. Langevin \cite{Langev21} made  clear that the Sagnac effect can be explained inside special relativity and, since the effect  involves a non-inertial reference frame, he suggested its compatibility with general relativity too.	 
Actually the Sagnac effect was considered in the context of general relativity only several years later by A. Ashtekar and A. Magnon \cite{AshMag75},
who introduced the notion of a {\em Sagnac tube} as a timelike surface $\mu$ in the spacetime $(M,g)$. Under the assumption that the Sagnac tube is foliated by the flow lines of a timelike Killing vector field in $\mu$,  they computed the {\em Sagnac shift} $\Delta \tau$, i.e. the difference of the values of the proper time of an observer (whose world line coincides with one of the Killing field flow line as an unparameterized curve) at the arrival points of two future-pointing  null curves   winding round the tube. 	 
Although, by the point of view of an experimental apparatus, it is completely reasonable   that the null curves considered  are not necessarily lightlike geodesics of the spacetime $(M,g)$, it would be desirable to analyze the possibility to detect the effect with freely falling light rays.  We point out  in the next section that,  in general, this is not guaranteed, even  under the stronger assumption that the tube is inside a stationary region, because of  the non-reversibility of the Finsler metric describing the lightlike geodesic flow in a stationary spacetime. We consider then a simple situation  where two winding round future-pointing lightlike geodesics can be found, and we submit evidence for the relations of the effect with the topology of the spacetime. In particular a non-trivial topology of the spacetime can cause the effect also for a static local observer as it was already observed in \cite{Stache82}. Finally, we remark that the arguments above  extend to freely falling massive particles.

\section{Sagnac tube in a stationary region}
Let $(M,g)$ be a spacetime and $\mathcal D\subset M$ be an open region which is invariant by the flow of a timelike Killing vector field $K$. Let us assume for simplicity that $\mathcal D$ splits as $S\times \R$, with $S$ a spacelike hypersurface (with boundary), and the natural coordinate $t$ associated to $\R$ induces the vector field $\partial_t=K$.
Thus the metric $g$ in $D$ is given by 
\beq\label{standstat}
g=- \Lambda d t^2+\omega \otimes d t+dt\otimes \omega +g_0,
\eeq
where $g_0$ is the Riemmanian metric induced by $g$ on $S$, $\omega$ is the one-form on $S$ metrically equivalent to the vector field on $S$ obtained as  the pointwise $g$-orthogonal projection of $K_x$, $x\in S$,  on $T_xS$, and $\Lambda=-g(K,K)|_S>0$.
As it is well-known, the mixed term $\omega dt$ in the metric expression is at the base of the Sagnac effect. 
In fact, a lightlike vector $(\tau,v)\in TD$ has $\tau$-component, in dependence of $v\in TS$, which is a root of the equation $g\big((\tau,v),(\tau, v)\big)=0$:
\beq\label{tau}
\tau=\frac{\omega(v)}{\Lambda}\pm\left (\frac{g_0(v,v)}{\Lambda}+\frac{\omega^2(v)}{\Lambda^2}\right)^{1/2}.
\eeq
Notice that the positive root corresponds to a future-pointing lightlike vector while the negative one to a past-pointing  one.
Let us denote by $h$ the Riemannian metric on $S$ defined as 
\[h:=\frac{g_0}{\Lambda}+\frac{\omega}{\Lambda}\otimes \frac{\omega}{\Lambda}\]
The expressions in \eqref{tau}  define two positive Lagrangians on $TS$ as 
\[F_{\pm}(v):=\pm \omega(v)/\Lambda+\big(h(v,v)\big)^{1/2}\]
which are two Finsler metrics of Randers type, called {\em Fermat metrics} in \cite{CaJaMa11}\footnote{We point out  that in other references, as e.g. \cite{Perlic04}, the name Fermat metric has been attributed to the Riemannian metric $h$.}.

Notice that these two metrics are non-reversible, i.e. $F_\pm(v)\neq F_\pm(-v)$, and induce then two different asymmetric distances on $S$ that can be used by the observer $K/\sqrt{\Lambda}$ to determine, at least locally,  both the time and the radar distance  (see, e.g. \cite{Perlic08}) on $\mathcal D$. In fact, each geodesic $x\colon [a,b]\to S$ of $F_+$ (resp. $F_-$) parameterized with $\frac{g_0(\dot x,\dot x)}{\Lambda}+\frac{\omega^2(\dot x)}{\Lambda^2}=\mathrm{const.}$ lifts, up to translation by the flow of the Killing field $K$, to a unique future-pointing (resp. past-pointing) lightlike geodesic $\gamma$ of $(\mathcal D,g)$ given by $\gamma_+(s)=\left(\int_a^sF_+(\dot x(u))du, x(s)\right)$ (resp. $\gamma_-=\left(-\int_a^sF_-(\dot x(u))du, x(s)\right)$) (see \cite[Th. 4.1]{CaJaMa11}). 

\bere\label{nonspacelike}
The Lagrangians $F_\pm$ make sense whenever $h$ is Riemannian without necessarily assuming that $g_0$ is a Riemannian metric. Of course, the signature of $g_0$ is related to the causal character of the hypersurface $S$ considered in the stationary splitting $S\times\R$. Notice however  that, if we assume that   $S$ is  timelike   or lightlike, then there exists a vector $v\in TS$ such that $\frac{1}{\Lambda}g_0(v,v)+\frac{1}{\Lambda^2}\omega^2(v)=h(v,v)\leq \frac{1}{\Lambda^2}\omega^2(v)$. This implies that the norm of $\omega/\Lambda$ w.r.t. $h$ is not strictly less than $1$. As a consequence, the Lagrangians $F_\pm$ are not Finsler in the classical sense (i.e. they are not positive,  and they do not have strongly convex indicatrices). Nevertheless, the critical curves of their action functionals (they satisfy the Euler-Lagrange equation \eqref{lorentzeq} below) can  still be lifted to, respectively,  future-pointing and past-pointing  null curves of the spacetime and, since $F_\pm(v)\neq F_\pm(-v)$,  they are in general not invariant by orientation reversing reparametrizations. Thus,  also  in this more general situation,  the arguments below related to the Sagnac effect  hold unchanged.  
\ere

Let us denote by $\ell_{F_\pm}$ the length functional associated to the Fermat metrics $F_\pm$, i.e. $\ell_{F_\pm}(\sigma):=\int_a^bF_\pm(\dot\sigma)ds$, where $\sigma:[a,b]\to S$, $\sigma=\sigma(s)$,   is a curve on $S$. Let us assume that there exists a (non-constant) geodesic  loop or a periodic geodesic  $x:[a,b]\to S$ of $F_+$ which remains a geodesic also when parameterized in the opposite direction. Then the Sagnac shift for the observer $K/\sqrt{\Lambda}$ along its world line passing through $x(\bar s)$, $\bar s\in[a,b]$,\footnote{Notice that if $x$ is a reversible geodesic loop based at $x(a)=x(b)$ and $\bar  s\in (a,b)$ then the two future-pointing lightlike curves  defined  by $x$ are piecewise lightlike geodesics.} is given  by
\beq\label{shift}|\Delta \tau| =\frac{1}{\sqrt{\Lambda(x(\bar s))}}\big|\ell_{F_+}(x)-\ell_{F_-}(x)\big|=\frac{2}{\sqrt{\Lambda(x(\bar s))}} \left|\int_x \omega/\Lambda\right|\eeq
that, by Stokes' theorem,  coincides with the value of the surface integral 
\[\frac{2}{\sqrt{\Lambda(x(\bar s))}}\left|\int_A d\left(\omega/\Lambda\right)\right| \]
if there exists a surface $A\subset S$ spanning $x$ (see also \cite{AshMag75,Frauen18}).

The above reasoning is based on the fact that $x$ remains a geodesic of $F$ when it is parameterized in the opposite direction. In general, the non-reversibility of the Fermat metric  implies  that this might be not the case.\footnote{Of course, if $x=x(s)$ is a geodesic of $F_\pm$ then $\tilde x(s)=x(a+b-s)$ is a geodesic of $F_{\mp}$.}
Indeed, if there exists a loop or a periodic geodesic $x:[a,b]\to S$ for the Fermat  metric $F_+$, we can  consider the Sagnac tube defined by $x$ and the flow lines of $K$ passing through the points in the image of  $x$ and then the future-pointing  lightlike geodesic $\gamma_+$ which connects, on this tube, the events  $(0,x(a))$ and $(\int_a^b F_+(\dot x)d s, x(a))$. Now, in general, the opposite curve of $x$ will be not a pregeodesic of $F_+$ and then there is no other distinct future-pointing lightlike geodesic in the Sagnac tube.
It is not difficult to construct examples where this can happen. It is enough to consider a Randers space $S$ with a non-reversible geodesic  and then to associate to it a standard stationary spacetime $\R\times S$ by  stationary-to-Randers correspondence \cite{CaJaSa11}. For example, consider the standard stationary spacetime in $\R\times D(0,2)$, where $D(0,2)\subset \R^2$ is the open disk centered in $0$ and having radius $2$, with the metric 
\[g=-dt^2+\omega\otimes  dt+dt\otimes \omega +   h   -\omega\otimes\omega,\]
where $  h  $ is the Euclidean metric in $\R^2$ and  $\omega$ is the one-form whose components are given by $\frac 1 2(y,-x)$, $(x,y)\in D(0,2)$.
It is immediate to see that the Fermat metric  $F_+$ associated to this spacetime is the Randers metric $F_+(v) = \sqrt{  h(v,v) }+ \omega (v)$, 
It can be shown that $F_+$   has closed geodesics given by all the circles of radius $1$ inside $D(0,2)$ but only if traversed anticlockwise \cite{Crampi05}. 

On the other hand,  it is possible that a closed geodesic of a Randers metric remains a geodesic also when parameterized in the opposite direction. This is for example  the case of a Katok metric on the sphere $\mathbb S^n$, i.e.  a  Randers metric  obtained via  Zermelo navigation   by considering as a wind  a Killing vector field on the sphere \cite[\S 4]{Robles07}. The orbits of the Killing vector field corresponding to a great circle on $S^n$ are geodesics of the Katok metric and when parameterized in the direction of the rotation have Finslerian length equal to $\frac{2\pi}{1+a}$ while in the opposite direction the length is    $\frac{2\pi}{1-a}$, where $0<a<1$ is a parameter associated to the rotation considered. Thus, from Eq. \eqref{shift}, the Sagnac effect, in the standard stationary spacetime associated to these data as above, is equal to
\[\Delta \tau=\frac{2\pi}{1-a}-\frac{2\pi}{1+a}=\frac{4\pi a}{1-a^2},\] 
independently of the world line of the observer  $\partial_t$ considered.
\section{Sagnac effect in a static spacetime with a  stationary splitting}
A case when any geodesic $x$ of $F_\pm$ remains a geodesic also when parameterized in the opposite direction is exactly when the one-form  $\omega/\Lambda$ is closed. 
Indeed, by using the Levi-Civita connection $\nabla$ of $h$ we obtain the Euler-Lagrange equation of the length functionals associated to $F_\pm$  as
\beq\label{lorentzeq}\nabla_{\dot x}\dot x=\pm \sqrt{h(\dot x,\dot x)}\hat\Omega(\dot x),\eeq
(see, e.g. \cite[Eq. (5)]{CaJaMa10b}) where $\hat\Omega:TS\to TS$ is the endomorphism $h$-metrically equivalent to $ d(\omega/\Lambda)$, when $x$ is parameterized with $h(\dot x, \dot x)=\mathrm{const.}$ 
Therefore, if $d(\omega/\Lambda)=0$ then $\hat\Omega=0$ and \eqref{lorentzeq} reduces to $\nabla_{\dot x}\dot x=0$ (i.e., in this case, $x$ is a geodesic of $F_\pm$ if and only if $x$ is a geodesic of $h$ and, moreover, if $x$ is a geodesic of $F_\pm$ then the opposite curve $\tilde x$ remains a geodesic of $F_\pm$). 

Thus, if   $S$ is not simply connected, the equation
$d(\omega/\Lambda)=0$ does not imply that $\omega/\Lambda$ is
exact; then the integral in \eqref{shift} is not in general equal to $0$.

Let us finally observe that if $\omega/\Lambda$ is closed then  $(D,g)$ is static with respect to the observer field $K/\sqrt{\Lambda}$ (see \cite[Def. 12.35]{One83}). In fact, let $(\bar t ,\bar p)\in \mathcal D$ and consider a neighborhood  $U\subset S$ of $\bar p$ such that a local primitive $f$ of $\omega/\Lambda$ is defined in $U$. By adding  a constant to $f$ we can assume that  $f(\bar p)=\bar t$. The graph $\mathcal G$ of $f$ is then an integral manifold of the orthogonal distribution defined by $K$  passing through $(\bar t,\bar p)$. Indeed, any vector $\zeta\in T\mathcal G$ is given by $(df(\xi), \xi)$ with $\xi\in TU$ and then $g(K,\zeta)=\omega(\xi)-\Lambda df(\xi)=0$.

Spacetimes satisfying the above assumptions can be found inside the class of  stationary cylindrical vacuum spacetimes (see, e.g., \cite[\S 6.1]{BrSaWa20} and the discussion in \cite{Stache82} about their existence as physical meaningful  solutions of the Einstein field equations).

\bere
It makes sense to consider the more general case  where the Killing field $K$ becomes  null in some embedded submanifold  of $\mathcal D$ (a Killing horizon). In this case it is still possible to introduce a positive homogeneous Lagrangian on $S$ related to future-pointing lightlike vectors in $(\mathcal D, g)$ (see {\em Randers-Kropina metrics} introduced  in \cite{CaJaSa14}). Anyway at the Killing horizon, all future-pointing lightlike vectors (except for the vectors collinear to $K$) can point  outside   $\mathcal D$ making impossible, without further assumptions,  to find a  future-pointing null curve  with endpoints  at a flow line of $K$ and that  remains in $\mathcal D$ (apart from the same flow line).  In particular, the existence of a closed geodesic for a Randers-Kropina metric on a compact manifold  is open, in general, even in the case when the Killing field is null everywhere (see \cite{CaJaSa14,CaGiMS21}). 
On the other hand, if $K$ is timelike everywhere, several results about existence and multiplicity of lightlike future-pointing  geodesics that project on  geometrically distinct closed geodesics of the Fermat metric $F_+$ are available when $S$ is a compact manifold (see \cite{BilJav08}). Such closed geodesics might be sources of the Sagnac effect with freely falling light rays.   
\ere
\section{Finslerian description of the Sagnac effect for massive particles}
It has been observed that the Sagnac shift  is {\em universal} in the sense that it does not depend on the physical nature of the two interfering beams: light rays, electromagnetic waves, neutron beams, etc. (see \cite{RizRug03}  and the references therein).
This universal property holds in the setting considered above when considering   freely falling particles.   
Indeed, future-pointing timelike geodesics $\gamma$   parameterized with proper time   in $\mathcal D$ can  be obtained from  the geodesics of the Fermat metric $\tilde F_+$ associated to the one-dimensional higher standard stationary spacetime $\mathcal D\times\R$ with the product metric
$\tilde g:=g\oplus d s^2$,  $g$ as in \eqref{standstat} and $s$ the natural coordinate on the added factor $\R$.
Since  $\partial_s$ is a Killing vector field for the metric  $\tilde g$, geodesics $\varsigma $ in $\mathcal D\times\R$ have to satisfy the conservation law 
$\tilde g(\dot \varsigma,\partial_s)=\mathrm{const.}$,
which  implies that the $s$-component of  a  geodesic
is an affine function.
Moreover, the projection $\gamma$ on $\mathcal D$ of  $\varsigma$ is a geodesic for $(\mathcal D,g)$.
In particular  lightlike geodesics $\varsigma=(\gamma,u)$ for the metric $\tilde g$ satisfy the  equation $g(\dot\gamma,\dot\gamma)=-\dot u^2=\mathrm{const.}$
Thus,  a timelike geodesic $\gamma$ in $(\mathcal D,g)$   parameterized with proper time   is the projection of a 
lightlike geodesic in $(\mathcal D\times\R,\tilde g)$ whose $s$-component $u$ has constant derivative equal to $  \pm 1 $. 
The  Fermat metrics $\tilde F_\pm$ associated to $\mathcal D\times\R$ are now the two Randers metrics on $S\times\R$ defined by
\[
\tilde F_\pm (v,\nu))=\pm \omega(v)/\Lambda+\Big(\tilde h\big((v,\nu),(v,\nu)\big)\Big)^{1/2},
\]
for all $(v,\nu))\in TS\times\R$, where $\tilde h$ is the Riemannian metric on $S\times \R$ defined as 
\[\tilde h:=\frac{g_0\oplus ds^2}{\Lambda}+\frac{\omega}{\Lambda}\otimes \frac{\omega}{\Lambda},\]
(see \cite[\S 4.3]{CaJaMa11}).
Thus, if $\gamma_i: [a,b]\to \mathcal D$, $i=1,2$ are two timelike geodesics of $(\mathcal D,g)$,   and  which project onto the same loop or closed curve $x$ in $S$ traversed in opposite directions, then 
the Sagnac shift, measured by the observer $\partial_t/\sqrt{\Lambda}$ along its world line passing through $x(\bar s)$, $\bar s\in [a,b]$,  is  given by
\[|\Delta \tau| =\frac{1}{\sqrt{\Lambda(x(\bar s))}}\big|\ell_{\tilde F_+}(x,u)-\ell_{\tilde F_-}(x,u)\big|=\frac{2}{\sqrt{\Lambda(x(\bar s))}} \left|\int_x \omega/\Lambda\right|\]
as in \eqref{shift}.



\section*{Acknowledgments}   The authors would like to express their gratitude to an anonymous referee for her/his comments and, in particular, for an observation that has been taken into account in Remark~\ref{nonspacelike} and for the suggestion  to consider, in the last section, freely falling particles parameterized with  proper time instead of  any fixed affine parameter.



\begin{thebibliography}{10}
	\providecommand{\url}[1]{{#1}}
	\providecommand{\urlprefix}{URL }
	\expandafter\ifx\csname urlstyle\endcsname\relax
	\providecommand{\doi}[1]{DOI \discretionary{}{}{}#1}\else
	\providecommand{\doi}{DOI \discretionary{}{}{}\begingroup
		\urlstyle{rm}\Url}\fi
	
	\bibitem{Ashby03}
	Ashby, N.: Relativity in the global positioning system. Living Rev. Relativity \textbf{6}, 1 (2003).
	\newblock \doi{10.12942/lrr-2003-1}
	
	\bibitem{Seller04}
	Selleri, F.: Sagnac effects: end of the mystery. In Rizzi, G., Ruggiero, M.L. (Eds.) Relativity in Rotating Frames. Relativistic Physics in
	Rotating Reference Frames, pp.
	57--77.  Kluwer Academic Publishers, Dordrecht (2004)
	
	\bibitem{Langev21}
	Langevin, P.: Sur la th\'eorie de relativit\'e et l'exp\'erience de {M}. {S}agnac. Comptes Rendus  des S\'eances de l'Acad\'emie des
	Sciences, Paris, \textbf{173}, 831--834 (1921)
	
	\bibitem{AshMag75}
	Ashtekar, A., Magnon, A.: The {S}agnac effect in general relativity. J. Math. Phys. \textbf{16}, 341--344
	(1975).
	\newblock \doi{10.1063/1.522521}
	
	\bibitem{Stache82}
	Stachel, J.: Globally stationary but locally static space-times: a gravitational
	analog of the {A}haronov-{B}ohm effect. Phys. Review D. \textbf{26}, 1281--1290 (1982).
	\newblock \doi{10.1103/PhysRevD.26.1281}
	
	\bibitem{CaJaMa11}
	Caponio, E., Javaloyes, M.A., Masiello, A.: On the energy functional on {F}insler manifolds and applications to stationary spacetimes. Math. Ann. \textbf{351},
	365--392 (2011).
	\newblock \doi{10.1007/s00208-010-0602-7}
	
	\bibitem{Perlic04}
	Perlick, V.:  Gravitational lensing from a spacetime perspective. Living Rev. Relativity \textbf{7}, 9 (2004).
	\newblock \urlprefix\url{http://www.livingreviews.org/lrr-2004-9}
	
	\bibitem{Perlic08}
	Perlick, V.: On the Radar Method in General-Relativistic Spacetimes. In Dittus, H.,  Lammerzahl, C., Turyshev, S.G. (Eds) Lasers, Clocks and Drag-Free Control: Exploration of Relativistic Gravity in Space, pp. 131--152. Springer, Berlin, Heidelberg, (2008).
	\newblock \doi{10.1007/978-3-540-34377-6_5}.
	\newblock \urlprefix\url{https://doi.org/10.1007/978-3-540-34377-6_5}
	
	\bibitem{Frauen18}
	Frauendiener, J.: Notes on the {S}agnac effect in general relativity. Gen. Rel. Grav. \textbf{50}, 
	147 (2018).
	\newblock \doi{10.1007/s10714-018-2470-5}
	
	\bibitem{CaJaSa11}
	Caponio, E., Javaloyes, M.A., S\'{a}nchez, M.:  On the interplay between {L}orentzian causality and {F}insler metrics of {R}anders type. Rev. Mat. Iberoam. \textbf{27}, 919--952 (2011).
	\newblock \doi{10.4171/RMI/658}
	
	\bibitem{Crampi05}
	Crampin, M.: Randers spaces with reversible geodesics. Publ. Math. Debrecen \textbf{67}, 401--409 (2005)
	
	\bibitem{Robles07}
	Robles, C.: Geodesics in {R}anders spaces of constant curvature. Trans. Amer. Math. Soc. \textbf{359}, 1633--1651 (2007).
	\newblock \doi{10.1090/S0002-9947-06-04051-7}
	\bibitem{CaJaMa10b}
	Caponio, E., Javaloyes, M.A., ~Masiello, A.:  Finsler geodesics in the presence of a convex function and their applications. J. Phys. A.  \textbf{43}, 135207, 15 (2010).
	\newblock \doi{10.1088/1751-8113/43/13/135207}
	
	\bibitem{One83}
	O'Neill, B.: Semi-{R}iemannian Geometry. Academic Press Inc., New York, (1983)
	
	\bibitem{BrSaWa20}
	Bronnikov, K.A.,  Santos, N.O., Wang, A.: Cylindrical systems in general relativity. Classical  Quantum Gravity
	\textbf{37}, 113002 (2020).
	\newblock \doi{10.1088/1361-6382/ab7bba}
	
	
	\bibitem{CaJaSa14}
	Caponio, E.,  Javaloyes,  M.A., S\'anchez, M.: Wind 
	Finslerian structures: from Zermelo's navigation to the causality of spacetimes. Available at 
	\urlprefix\url{http://arxiv.org/pdf/1407.5494v5},  (2014)
	
	\bibitem{CaGiMS21}
	Caponio, E. Giannoni, F., Masiello, A., Suhr, S.: Connecting and closed geodesics of a Kropina metric. Adv. Nonlinear Stud. \textbf{21}, 683--695 (2021)
	(2021).
	\newblock \doi{doi:10.1515/ans-2021-2133}.
	\newblock \urlprefix\url{https://doi.org/10.1515/ans-2021-2133}
	
	\bibitem{BilJav08}
	Biliotti, L., Javaloyes, M.~A.: $t$-periodic light rays in conformally stationary spacetimes via {F}insler geometry. Houston J. Math. \textbf{37}, 127--146 (2011)
	
	\bibitem{RizRug03}
	Rizzi, G., Ruggiero, M.L.: A direct kinematical derivation of the relativistic Sagnac effect for
	light or matter beams. Gen. Rel. Grav. \textbf{35},  2129--2136  (2003).
	\newblock \doi{10.1023/A:1027345505786}
	
\end{thebibliography}

\end{document}